\documentclass[12pt]{article}
\input epsf.tex
\newcommand{\beq}{\begin{equation}}
\newcommand{\eeq}{\end{equation}}
\newcommand{\beqa}{\begin{eqnarray}}
\newcommand{\eeqa}{\end{eqnarray}}
\newcommand{\beqar}{\begin{eqnarray*}}
\newcommand{\eeqar}{\end{eqnarray*}}

\newcommand{\eps}{\epsilon}

\newcommand{\inn}{\!\cdot\!}

\newcommand{\la}{\lambda}

\newcommand{\z}{\zeta}

\newcommand{\eg}{{\it e.g.,}\ }
\newcommand{\ie}{{\it i.e.,}\ }
\newcommand{\labell}[1]{\label{#1}} %{\label{#1}} %
\newcommand{\reef}[1]{(\ref{#1})}

\newcommand\veps{\varepsilon}

\newcommand\cJ{{\cal J}}
\newcommand\cI{{\cal I}}

\newcommand\cR{{\cal R}}

\newcommand\tG{{\widetilde G}}

\newcommand\tphi{{\widetilde \phi}}

\newcommand\ta{{\tilde a}}
\newcommand\tb{{\tilde b}}

\newcommand\te{{\tilde e}}
\newcommand\tk{{\tilde k}}

\newcommand\tf{{\tilde f}}
\newcommand\tl{{\tilde l}}

\newcommand\tB{{\widetilde B}}

\newcommand\tC{{\widetilde C}}

\newcommand\tr{{\rm tr}}
\begin{document}

\begin{titlepage}

\begin{center}

%\fbox{DRAFT: \today}

%{} \hfill     XX-X-X, \, MIFP-XX-XX

\vskip 6 cm
\bf{{\LARGE T-duality of anomalous Chern-Simons  couplings }}\\
\vskip 1.25 cm
 Mohammad R. Garousi\footnote{garousi@ferdowsi.um.ac.ir}  \\
 
\vskip 1 cm
{{Department of Physics, Ferdowsi University of Mashhad\\}{P.O. Box 1436, Mashhad, Iran}\\}
\vskip .1 cm
%and
%\vskip .1 cm
%{{School of Physics, Institute for Research in Fundamental Sciences (IPM)\\}{P.O. Box 19395-5531, Tehran, Iran}\\}

\end{center}

\vskip 0.5 cm

\begin{abstract}
\baselineskip=18pt

It is known  that the anomalous D$_p$-brane  Chern-Simons couplings  are not consistent with the
standard rules of T-duality. Using compatibility of these couplings with the linear T-duality transformations,   the B-field gauge transformations   and  the general coordinate transformations as  guiding principles we find new couplings at order $O(\alpha'^2)$ for ${\cal C}^{(p-3)}$, ${\cal C}^{(p-1)}$ , ${\cal C}^{(p+1)}$ and  ${\cal C}^{(p+3)}$. %, ${\cal C}^{(p+3)}$ and for ${\cal C}^{(p+5)}$. 
%The couplings involving the R-R field strength ${\cal F}^{(p-2)}$ are invariant under T-duality. They are covariant at the level of two NS-NS fields, and are invariant under the   $B$-field  gauge transformations. The couplings  involving  ${\cal F}^{(p)}$ and ${\cal F}^{(p+2)}$, however,  have  some extra non-covariant terms which may  be combined with the corresponding couplings at the level of  one NS-NS field  to make them  covariant at the level of two NS-NS fields. 

\end{abstract}

\vskip 0.5 cm


\end{titlepage}

\section{Introduction }
The dynamics of the D-branes of type II superstring theories is well-approximated by the effective world-volume field theory  which consists of the  Dirac-Born-Infeld (DBI) and the Chern-Simons (CS) actions. 
The DBI action  describes the dynamics of the brane in the presence of  NS-NS background fields. For constant  fields,  this action can be found by requiring its  consistency with the  nonlinear T-duality \cite{Leigh:1989jq,Bachas:1995kx}, \ie
\beqa
S_{DBI}&=&-T_p\int d^{p+1}x\,e^{-\phi}\sqrt{-\det\left(G_{ab}+B_{ab}+2\pi\alpha'f_{ab}\right)}\labell{DBI}
\eeqa
where $G_{ab}$ and $B_{ab}$ are  the pull-back of the bulk fields $G_{\mu\nu}$ and $B_{\mu\nu}$ onto the world-volume of D-brane\footnote{Our index convention is that the Greek letters  $(\mu,\nu,\cdots)$ are  the indices of the space-time coordinates, the Latin letters $(a,d,c,\cdots)$ are the world-volume indices and the letters $(i,j,k,\cdots)$ are the normal bundle indices.}. The curvature corrections to this action have been found in \cite{Bachas:1999um} by requiring the consistency of the effective action with the $O(\alpha'^2)$ terms of the corresponding disk-level scattering amplitude \cite{Garousi:1996ad,Hashimoto:1996kf}.  The B-field corrections  at  this order   have been found in \cite{Garousi:2009dj} by requiring the consistency of the curvature couplings with the linear T-duality transformations.

 On the other hand,  the CS part describes the coupling of D-branes to the R-R potential. For constant  fields it is given by \cite{Polchinski:1995mt,Douglas:1995bn}
\beqa
S_{CS}&=&T_{p}\int_{M^{p+1}}e^{B}C\labell{CS2}
\eeqa
where $M^{p+1}$ represents the world volume of the D$_p$-brane, $C$ is meant to represent a sum over all appropriate R-R  forms  and the multiplication rule is the wedge product. The abelian gauge field can be added to the action as $B\rightarrow B+2\pi\alpha'f$. Curvature correction to this action has been found in \cite{Green:1996dd,Cheung:1997az,Minasian:1997mm}
 by requiring that the chiral anomaly on the world volume of intersecting D-branes (I-brane) cancels  the anomalous variation of the CS action. This correction is
\beqa
S_{CS}&=&T_{p}\int_{M^{p+1}}{\cal C}\left(\frac{{\cal A}(4\pi^2\alpha'R_T)}{{\cal A}(4\pi^2\alpha'R_N)}\right)^{1/2}\labell{CS}
\eeqa
where ${\cal C}=e^BC$ and ${\cal A}(R_{T,N})$ is the Dirac roof genus of the tangent and normal bundle curvatures respectively,
\beqa
\sqrt{\frac{{\cal A}(4\pi^2\alpha'R_T)}{{\cal A}(4\pi^2\alpha'R_N)}}&=&1+\frac{\pi^2\alpha'^2}{24}(\tr R_T^2-\tr R_N^2)+\cdots \labell{roof}
\eeqa
For totally-geodesic embeddings of the world-volume in the ambient spacetime,   ${\rm R}_{T,N}$ are the pull-back curvature 2-forms of the tangent and normal bundles respectively (see the appendix in ref.
\cite{Bachas:1999um} for more details). 

It was shown in \cite{Garousi:2010ki} that at order $O(\alpha'^2)$ the CS action \reef{CS}  must  include additional linear couplings to the NS-NS fields.  These couplings were found  by studying the S-matrix element of one R-R and one NS-NS vertex operator  at order $O(\alpha'^2)$ \cite{Garousi:1996ad}. In the string frame, they  take the form \cite{Garousi:2010ki}\footnote{Using the on-shell relations, the standard definition of the curvature tensor $\hat{R}_{ij}$  has been changed in \cite{Garousi:2010ki} to $\hat{R}_{ij}\equiv \frac{1}{2}(R_{ia}{}^a{}_j-R_{ik}{}^k{}_j)$. With this tensor the coupling $F^{(p+2)}_{a_0\cdots a_pj,i}\hat{R}^{ij}$ is then invariant under linear T-duality \cite{Garousi:2010ki}. If one uses the standard definition $\hat{R}_{ij}\equiv R_{ia}{}^a{}_j$, then the second term in the second line of \reef{LTdual} can be written at the linear order as $F^{(p+2)}_{a_0\cdots a_p}{}^{ j,i}(h_{ij,aa}+h_{aa,ij}-h_{ia,aj}-h_{ja,ai}-2\phi_{,ij})/2(p+1)$ where $h$ is the metric perturbation. Under T-duality along the world volume direction  $y$, the RR factor $F^{(p+2)}_{a_0\cdots a_p}{}^{ j,i}/(p+1)$ which includes the Killing index $y$, transforms to $F^{(p+1)}_{a_0\cdots a_{p-1}}{}^{ j,i}$. The latter, however,  does not include the Killing index. Hence, the indices $i,j$ in the T-dual theory do not include the Killing index $y$. Using this observation, one can easily verify  that the metric/dilaton factor $(h_{ij,aa}+h_{aa,ij}-h_{ia,aj}-h_{ja,ai}-2\phi_{,ij})$ is invariant under the linear T-duality. Hence, the second term in the second line of \reef{LTdual} is invariant under the T-duality.}:
\beqa
S_{CS}&\!\!\!\!\!\supset\!\!\!\!&\pi^2\alpha'^2T_p\int d^{p+1}x\,\eps^{a_0\cdots a_p}\left(\frac{1}{2!(p-1)!}[{ F}^{(p)}_{ia_2\cdots a_p,a}H_{a_0a_1}{}^{a,i}-{ F}^{(p)}_{aa_2\cdots a_p,i}H_{a_0a_1}{}^{i,a}]\right.\nonumber\\
&&\left.\qquad\qquad\qquad\qquad+\frac{2}{p!}[\frac{1}{2!}{ F}^{(p+2)}_{ia_1\cdots a_pj,a}\cR^a{}_{a_0}{}^{ij}-\frac{1}{p+1}{ F}^{(p+2)}_{a_0\cdots a_pj,i}(\hat{\cR}^{ij}-\phi\,^{,ij})]\right.\nonumber\\
&&\left.\qquad\qquad\qquad\qquad-\frac{1}{3!(p+1)!}{ F}^{(p+4)}_{ia_0\cdots a_pjk,a}H^{ijk,a}\right)\labell{LTdual}
\eeqa
where ${\cal R}$ is the linearized Riemann curvature tensor of the background metric,  $F^{(n)}=dC^{(n-1)}$, and  commas are used to denote partial differentiation. Since these couplings have been found by  the S-matrix method, there is an  on-shell ambiguity in defining these terms \cite{AAT,AAT1}.  
The above couplings   are consistent  with the T-duality transformations at a linearized level and are invariant under the $B$-field gauge transformations. In particular, the sum of the second term in the first line and the last two terms in the second line form a T-duality invariant set of terms, and the remaining terms  form another T-duality invariant set. We call each of these  a T-dual multiplet. 

One may extend  \reef{LTdual}  to the nonlinear couplings by replacing $C$ with ${\cal C}=e^BC$ and by replacing the ordinary derivatives with their covariant counter parts. In fact the first replacement is required for  consistency of the above couplings with the nonlinear T-duality transformations \cite{Garousi:1996ad}. When the R-R potential carries one transverse index, this replacement produces the following couplings for $C^{(p-3)}$ :   
\beqa
&&\frac{\pi^2\alpha'^2T_{p}}{2!(p-4)!}\int d^{p+1}x\,\eps^{a_0a_1\cdots a_{p}}\bigg[\left(\frac{1}{2!}C^{(p-3)}_{ia_2\cdots a_{p-3},a_{p-2}}B_{a_{p-1}a_{p}}
-\frac{1}{3!}C^{(p-3)}_{a_2\cdots a_{p-3}i}H_{a_{p-2}a_{p-1}a_{p}}\right)_{,a}\,H_{a_0a_1}{}^{a,i}\nonumber
\eeqa
The first  term breaks the B-field gauge symmetry. However,  it can be restored by the standard replacement of $B_{a_{p-1}a_{p}}$ with $(B_{a_{p-1}a_{p}}+2\pi\alpha'f_{a_{p-1}a_{p}})$. It has been shown in \cite{Garousi:2010bm} that the S-matrix element of one R-R potential  and two B-field vertex operators  reproduce exactly the above couplings. When the R-R potential carries only the world volume indices,  the above replacement does not restore the gauge symmetry in many terms. The non-gauge invariant terms, however,  are invariant under the linear T-duality at the level of two B-fields, so it is consistent with the linear T-duality to remove them. On the other hand,  the S-matrix calculations produce only the gauge invariant couplings \cite{Garousi:2011ut}.

It has been pointed out in \cite{Myers:1999ps} that the anomalous CS couplings \reef{CS} must be incomplete for non-constant B-field as they are not compatible with the T-duality. T-duality exchanges the components of the metric and  the B-field whereas  the couplings \reef{CS} involve only the metric through the curvature terms. A systematic approach  for including the B-field in a theory might be  provided by  the `double field theory` formalism in which the  fields depend both on the usual spacetime coordinates and on the winding coordinates \cite{Hull:2009mi}. In this paper, however,  we use the method that was used in \cite{Myers:1999ps} to find the Myers terms in the non-abelian CS action at order $O(\alpha'^0)$. That is, we add new couplings to the CS action at order $O(\alpha'^2)$ to make it  compatible with  T-duality.  The T-dual multiplet  that we  find includes the R-R potentials $C^{(p-3)},\, C^{(p-1)}$ and $C^{(p+1)}$.  These couplings have   been also found in \cite{Becker:2010ij}. They are, however,  neither covariant nor invariant under the B-field gauge transformations. 

The disk-level S-matrix element of one R-R potential $C^{(p-3)}$ and two B-field vertex operators produces not only  the $C^{(p-3)}$ component of the above T-dual multiplet, but also   produces  some other contact terms as well as massless poles at order $O(\alpha'^2)$ \cite{Garousi:2011ut}.  Consistency of the amplitude  with  linear T-duality then requires one to extend the latter contributions   to  contact-term and  the massless-pole T-dual multiplets. More generally, one may extend the S-matrix element to a set of S-matrix elements which are invariant under the linear T-duality transformations. We call this set  the S-matrix T-dual multiplet. 

Having both the contact-term  as well as  the massless-pole T-dual multiplets at order $O(\alpha'^2)$, it raises the question of how they come together  to produce explicit covariant/gauge-invariant results. One may expect that these  multiplets can be combined separately to become covariant/gauge-invariant. However, as we will  show the S-matrix calculation indicates that some of the terms in a contact-term  multiplet combine with the massless-pole  multiplets to produce the  covariant and gauge-invariant results.   We will show that such terms must be proportional to the Mandelstam variables. This phenomenon does not appear in the T-dual multiplets in \reef{LTdual} because the S-matrix element of two closed string vertex operators at order $O(\alpha'^2)$ has only contact terms \cite{Garousi:1996ad}.

The outline of the paper is as follows: We begin in section 2 by reviewing the T-duality transformations and the method for finding the  T-dual completion of a coupling.   In section 3.1, we show that the standard CS coupling \reef{CS}  is not consistent with the linear T-duality transformations and add new couplings  at order $O(\alpha'^2)$ to find its corresponding  T-dual multiplet.  The $C^{(p-3)}$ component of this CS multiplet, however, is not  invariant under the B-field gauge transformations. In sections 3.2, by adding another  T-dual multiplet, we write the $C^{(p-3)}$ component of the combined multiplet in  a T-dual and gauge-invariant form (see eq.\reef{Tf41new}).  In section 3.3, we argue that the contact terms in a S-matrix T-dual multiplet  which are proportional to the Mandelstam variables,  may  combine with the massless poles   to produce covariant and gauge-invariant results. Since we are not considering the massless poles of the S-matrix  multiplet in this paper,  we will not attempt  to make such contact terms to be covariant/gauge-invariant. Adding three contact-term T-dual multiplets to the CS multiplet, we then write the $C^{(p-1)}$ component of these multiplets in a covariant and gauge-invariant form (see eq.\reef{new24}). In section 3.4, by adding one more T-dual multiplet to the list, we write the $C^{(p+1)}$ components in a covariant and gauge-invariant form (see eq.\reef{new25}). Finally, we show in section 3.5 that the $C^{(p+3)}$ components of the above multiplets are covariant and gauge-invariant  (see eq.\reef{new26}). 

\section{T-duality}

The full set of nonlinear T-duality transformations for massless R-R and NS-NS fields have been found in \cite{TB,Meessen:1998qm,Bergshoeff:1995as,Bergshoeff:1996ui,Hassan:1999bv}. The nonlinear T-duality transformations of the  fields $C$ and $B$ are such that  the expression ${\cal C}=e^BC$  transforms linearly under  T-duality \cite{Taylor:1999pr}. When  the T-duality transformation acts along the Killing coordinate $y$,  the massless NS-NS fields and  ${\cal C}$  becom:
\beqa
e^{2\tphi}&=&\frac{e^{2\phi}}{G_{yy}}\nonumber\\
\tG_{yy}&=&\frac{1}{G_{yy}}\nonumber\\
\tG_{\mu y}&=&\frac{B_{\mu y}}{G_{yy}}\nonumber\\
\tG_{\mu\nu}&=&G_{\mu\nu}-\frac{G_{\mu y}G_{\nu y}-B_{\mu y}B_{\nu y}}{G_{yy}}\nonumber\\
\tB_{\mu y}&=&\frac{G_{\mu y}}{G_{yy}}\nonumber\\
\tB_{\mu\nu}&=&B_{\mu\nu}-\frac{B_{\mu y}G_{\nu y}-G_{\mu y}B_{\nu y}}{G_{yy}}\nonumber\\
{\cal \tC}^{(n)}_{\mu\cdots \nu y}&=&{\cal C}^{(n-1)}_{\mu\cdots \nu }\nonumber\\
{\cal \tC}^{(n)}_{\mu\cdots\nu}&=&{\cal C}^{(n+1)}_{\mu\cdots\nu y}\labell{Cy}
\eeqa
where $\mu,\nu\ne y$. In above transformation the metric is given in the string frame. If $y$ is identified on a circle of radius $R$, \ie $y\sim y+2\pi R$, then after T-duality the radius becomes $\tilde{R}=\alpha'/R$. The string coupling is also shifted as $\tilde{g}=g\sqrt{\alpha'}/R$.
We would like to study the consistency of  the CS couplings \reef{CS} with the linear T-duality transformations. Assuming that the NS-NS  fields are small perturbations around the flat space, the above  transformations take the following linear form:
\beqa
&&
\tilde{\phi}=\phi-\frac{1}{2}h_{yy},\,\tilde{h}_{yy}=-h_{yy},\, \tilde{h}_{\mu y}=B_{\mu y},\, \tilde{B}_{\mu y}=h_{\mu y},\,\tilde{h}_{\mu\nu}=h_{\mu\nu},\,\tilde{B}_{\mu\nu}=B_{\mu\nu}\nonumber\\
&&{\cal \tC}^{(n)}_{\mu\cdots \nu y}={\cal C}^{(n-1)}_{\mu\cdots \nu },\,\,\,{\cal \tC}^{(n)}_{\mu\cdots\nu}={\cal C}^{(n+1)}_{\mu\cdots\nu y}\labell{linear}
\eeqa

The strategy to find T-duality invariant couplings  is given in \cite{Garousi:2009dj}. Let us review it here.  
Suppose we are implementing T-duality along a world volume direction $y$ of a D$_p$-brane. First, we separate the  world-volume indices  along and orthogonal to  $y$ direction and then apply the  T-duality transformations.  The  orthogonal indices  are  the complete world-volume indices   of the  T-dual D$_{p-1}$-brane. However,  $y$  in the T-dual theory, which is a normal bundle index, is not  complete. 
%One must then include some other terms in the original theory to have totally completed indices in the T-dual theory. 
On the other hand, the normal bundle indices of  the original theory  are not complete in the T-dual D$_{p-1}$-brane. They do not include the $y$ index.  In a T-duality invariant theory,  $y$   must be combined  with the incomplete normal bundle indices  to make them complete.  If a theory is not invariant under the T-duality,  one should then add  new terms  to it  to have   the complete indices in the T-dual theory. In this way one makes the theory to be T-duality invariant by adding new couplings.

One may also  implement T-duality along a transverse direction $y$ of a D$_p$-brane. In this case, we  separate the transverse indices along and orthogonal to $y$ direction and then apply the  T-duality transformations. The latter indices  are   complete in the dual D$_{p+1}$-brane. However, the  complete world-volume indices of the original D$_p$-brane    are not complete in the dual D$_{p+1}$-brane. They must  include the $y$ index to be complete.  In a T-duality invariant theory,  $y$,  which is a world-volume index in the dual theory, must be combined with the incomplete world-volume indices  of the dual D$_{p+1}$-brane to become  complete. 

Let us apply the above method to the DBI action. Expansion of the DBI action \reef{DBI} produces the following terms at order $O(\alpha'^0)$:
\beqa
S_{DBI}=-T_p\int d^{p+1}x\bigg[1-\phi+\frac{1}{2}h_a{}^a
+\frac{1}{8}(h_a{}^a)^2-\frac{1}{4}h_a{}^b h_b{}^a-\frac{1}{4}B_a{}^b B_b{}^a+\frac{1}{2}\phi^2-\frac{1}{2}\phi h_a{}^a+\cdots\bigg]\nonumber
\eeqa
where  we  have considered perturbations around flat space. The metric takes the form  $G_{\mu\nu}=\eta_{\mu\nu}+h_{\mu\nu}$ where $h_{\mu\nu}$ is a small perturbation. We want to implement T-duality along a world volume direction. So  we write the linear terms above in the following form:
\beqa
-\phi+\frac{1}{2}h_a{}^a&=& -\phi+\frac{1}{2}h_\ta{}^\ta+\frac{1}{2}h_{yy}\nonumber
\eeqa
where the world volume index $\ta$ does not include $y$. Under the linear T-duality transformations \reef{linear},  it transforms to $-\phi+\frac{1}{2}h_\ta{}^\ta$. Since there is no incomplete index, one concludes that the linear terms in the DBI action are invariant under the linear T-duality transformations. Doing the same steps, one finds that the quadratic terms  transform under the linear T-duality transformations as
\beqa
\frac{1}{8}(h_\ta{}^\ta)^2-\frac{1}{4}(h_{yy})^2-\frac{1}{4}h_\ta{}^\tb h_\tb{}^\ta-\frac{1}{4}B_\ta{}^\tb B_\tb{}^\ta+\frac{1}{2}h_\ta{}^y h_y{}^\ta+\frac{1}{2}B_\ta{}^y B_y{}^\ta+\frac{1}{2}\phi^2-\frac{1}{2}h_\ta{}^\ta\phi\labell{linear1}
\eeqa
This expression includes terms with the  $y$ index. However, one should not conclude that the quadratic  terms are not invariant under the T-duality transformations. One has to add the nonlinear T-duality transformations of the linear terms $-\phi+h_a{}^a/2$, which include the following quadratic terms:
\beqa
\frac{1}{4}(h_{yy})^2-\frac{1}{2}h_\ta{}^y h_y{}^\ta-\frac{1}{2}B_\ta{}^y B_y{}^\ta \,, \nonumber
\eeqa
to the above couplings. This will cancel the terms in \reef{linear1} which have $y$ index. Hence, according to our expectations, the quadratic order terms in the DBI action are invariant under the T-duality transformations.

\section{New Couplings}

It is known  that  the anomalous CS couplings of D-branes to space-time curvature are incomplete, as they are inconsistent with T-duality. We will construct a form of the couplings which are consistent with the linear T-duality. We are interested in the $O(\alpha'^2)$ terms in \reef{roof}.  The world-volume curvature $R_T$ and the field strength $R_N$ are related to the pull-back of the space-time Riemann tensor and the second fundamental form though the Gauss-Codazzi equation:
\beqa
(R_T)_{abcd}&=&R_{abcd}+\delta_{ij}(\Omega^i_{ac}\Omega^j_{bd}-\Omega^i_{ad}\Omega^j_{bc})\nonumber\\
(R_N)_{ab}{}^{ij}&=&-R^{ij}{}_{ab}+G^{cd}(\Omega^i_{ac}\Omega^j_{bd}-\Omega^j_{ac}\Omega^i_{bd})\nonumber
\eeqa
where $\Omega$ is the second fundamental form (see the appendix in \cite{Bachas:1999um}). For totally-geodesic embedding,  $\Omega$ is zero. In the static gauge, that we are going to use in this paper, the second fundamental form is non-zero. Hence, at order $O(\alpha'^2)$ there are three different terms: Terms with two Riemann tensors, terms with one Riemann tensor and two fundamental forms, and terms with four  fundamental forms.  At the linearized level, the Riemann curvature tensor is the second derivative of the fluctuation of the space-time metric and the second fundamental form is the second derivative of  the massless transverse scalar fields on the D-brane.  In this paper we are interested in studying the T-duality transformation of the two Riemann curvature terms. Hence, we consider the following  CS couplings in \reef{CS}: 
\beqa
\frac{T_p}{2!2!(p-3)!}\int d^{p+1}x\epsilon^{a_0\cdots a_{p-4}abcd}{\cal C}^{(p-3)}_{a_0\cdots a_{p-4}}\bigg[R_{ab}{}^{ef}R_{cdfe}-R_{ab}{}^{kl}R_{cdlk}\bigg]\labell{Tf2}
\eeqa
where we have employed the static gauge. That is, first we have used the  spacetime diffeomorphisms to define the D$_p$-brane world-volume as $x^i=0$, where  $i=p+1,\cdots, 9$, and then with the world-volume diffeomorphisms, we matched the internal coordinates with the remaining spacetime coordinates on the surface: $\sigma^a=x^a$ for $a=0,1,\cdots, p$. We have also   ignored the pull-back operations, \ie   we work only with the restriction of the Riemann tensor to the appropriate subspace. 

To find the T-dual completion of the above couplings at the linearized level,  we will consider perturbation around flat space where the metric takes the form  $G_{\mu\nu}=\eta_{\mu\nu}+h_{\mu\nu}$, where $h_{\mu\nu}$ is a small perturbation. We denote the Riemann tensor to linear order in $h$ by $\cR_{\mu\nu\rho\lambda}$. This linear Riemann tensor is,
\beqa
\cR_{\mu\nu\rho\la}&=&\frac{1}{2}(h_{\mu\la,\nu\rho}+h_{\nu\rho,\mu\la}-h_{\mu\rho,\nu\la}-h_{\nu\la,\mu\rho})
\eeqa
 The coupling \reef{Tf2} at the linearized level is then 
\beqa
\frac{T_p}{2!(p-3)!}\int d^{p+1}x\epsilon^{a_0\cdots a_{p-4}abcd}{\cal C}^{(p-3)}_{a_0\cdots a_{p-4}}\bigg[h_{a}{}^f{}_{,b}{}^e(h_{ce,df}-h_{cf,de})-h_{a}{}^l{}_{,b}{}^k(h_{ck,dl}-h_{cl,dk})\bigg]\labell{Tf02}
\eeqa
The  indices that are contracted with the volume form are totally antisymmetric so we do not use the antisymmetric notation for them. The above couplings have been verified by the S-matrix element of one R-R and two graviton vertex operators in \cite{Craps:1998fn}. 
%The consistency of the above couplings with linear T-duality has been studied in \cite{Becker:2010ij}. This leads to new couplings for $C^{(p-3)}$ and two B-fields as well as new couplings for $C^{(p-1)}$ and $C^{(p+1)}$ \cite{Becker:2010ij}. Those couplings, however,  are not gauge invariant under the R-R gauge transformation.  To find the T-dual completion of the couplings \reef{Tf02} to be invariant under the R-R gauge transformation, we note that the expression in the big parenthesis   is an exact 4-form in the world volume, so the above coupling   can be rewritten as
%\beqa
%\frac{T_p}{2!(p-2)!}\int d^{p+1}x\epsilon^{a_0\cdots a_{p-3}abc}{\cal F}^{(p-2)}_{a_0\cdots a_{p-3}}\bigg(h_{a}{}^f{}_{,b}{}^e(h_{ce,f}-h_{cf,e})-h_{a}{}^l{}_{,b}{}^k(h_{ck,l}-h_{cl,k})\bigg)\labell{Tf023}
%\eeqa
%where we have used the relation $\epsilon^{a_0\cdots a_{p-4}d}{\cal C}^{(p-3)}_{a_0\cdots a_{p-4},d}=\epsilon^{a_0\cdots a_{p-4}d}{\cal F}^{(p-2)}_{a_0\cdots a_{p-4}d}/(p-2)$.  Consistency of this  expression  with T-duality requires one to add new terms in the big parenthesis. The new terms  would be  obviously  invariant under the ${\cal C}$-gauge transformation. Hence, 
We will examine the  expression \reef{Tf02}  under the linear T-duality transformations \reef{linear}, and find its corresponding T-dual multiplet. We  call this multiplet, which has  the Chern-Simons couplings in its first component, the Chern-Simons multiplet.  
%\beqa
%&&\epsilon^{a_0\cdots a_{p-8}abcda'b'c'd'}C^{p-7}_{a_0\cdots a_{p-8}}\bigg(h_{a}{}^f{}_{,b}{}^e(h_{ce,df}-h_{cf,de})-h_{a}{}^l{}_{,b}{}^k(h_{ck,dl}-h_{cl,dk})\bigg)\nonumber\\
%&&\qquad\qquad\qquad\qquad\times\bigg(h_{a'}{}^h{}_{,b'}{}^g(h_{c'g,d'h}-h_{c'h,d'g})-h_{a'}{}^n{}_{,b'}{}^m(h_{c'm,d'n}-h_{c'n,d'm})\bigg)\labell{Tf022}
%\eeqa
%\beqa
%\epsilon^{a_0\cdots a_{p-8}abcda'b'c'd'}C^{p-7}_{a_0\cdots a_{p-8}}\bigg(R_{ab}{}^{ef}R_{cdfg}R_{a'b'}{}^{gh}R_{c'd'he}-R_{ab}{}^{kl}R_{cdlm}R_{a'b'}{}^{mn}R_{c'd'nk}\bigg)\labell{Tf0222}
%\eeqa

\subsection{Chern-Simons multiplet }

We begin by implementing T-duality along a world volume direction of D$_p$-brane, which is denoted by $y$. From the contraction with the world volume form, one of the indices $a_0,\cdots, a_{p-4}$ of the R-R potential\footnote{In the literature , the R-R potential is $C$. However,  in this paper we always work with ${\cal C}=e^BC$ and  call it R-R potential.}or the indices  $a,c$ of the metric fluctuation  in \reef{Tf02} must include $y$. So there are two cases to consider: First when the R-R potential ${\cal C}^{(p-3)}$ carries the $y$ index and second when   the metric carries the $y$ index. In the former case, we  write \reef{Tf02} as
\beqa
\frac{T_p}{2!(p-4)!}\int d^{p+1}x\epsilon^{a_0\cdots a_{p-4}yabcc}{\cal C}^{(p-3)}_{a_0\cdots a_{p-5}y}\bigg[h_{a}{}^{f}{}_{,b}{}^{e}(h_{ce,df}-h_{cf,de})
-h_{a}{}^l{}_{,b}{}^k(h_{ck,dl}-h_{cl,dk})\bigg]\labell{22}
\eeqa
The indices $e$ and $f$ include the Killing coordinate $y$ which is a world volume coordinate. However, in the T-dual theory, $y$ is a transverse coordinate. To be able to use the T-duality transformation rules \reef{linear},  we separate $y$ from $e,f$. Hence, we write the above equation as  
\beqa
%&&2\epsilon^{a_0\cdots a_{p-5}yabcd}C^{(p-3)}_{a_0\cdots a_{p-5}y}\bigg(h_{a}{}^f{}_{,b}{}^e(h_{ce,f}-h_{cf,e})-h_{a}{}^l{}_{,b}{}^k(h_{ck,l}-h_{cl,k})\bigg)_{,d}=\labell{first0}\\
\frac{T_p}{2!(p-4)!}\int d^{p+1}x\epsilon^{a_0\cdots a_{p-4}yabcc}{\cal C}^{(p-3)}_{a_0\cdots a_{p-5}y}\bigg[h_{a}{}^{\tf}{}_{,b}{}^{\te}(h_{c\te,d\tf}-h_{c\tf,d\te})-h_a{}^y{}_{,b}{}^{\te}h_{cy,d\te}\nonumber\\
-h_{a}{}^l{}_{,b}{}^k(h_{ck,dl}-h_{cl,dk})\bigg]\nonumber
\eeqa
where the "tilde" over  the world volume indices $e,f$ means they do not include the Killing direction $y$. Now, the above equation  transforms under  \reef{linear} to the following couplings of D$_{p-1}$-brane:
\beqa
\frac{T_{p-1}}{2!(p-4)!}\int d^{p}x\epsilon^{a_0\cdots a_{p-4}abcd}{\cal C}^{(p-4)}_{a_0\cdots a_{p-5}}\bigg[h_{a}{}^{f}{}_{,b}{}^{e}(h_{ce,df}-h_{cf,de})-B_a{}^y{}_{,b}{}^{e}B_{cy,de}\nonumber\\
-h_{a}{}^{\tl}{}_{,b}{}^{\tk}(h_{c\tk,d\tl}-h_{c\tl,d\tk})\bigg]\labell{first00}
\eeqa
where we have used the fact that $T_p\sim1/g_s$ and the relation $2\pi\sqrt{\alpha'}T_p=T_{p-1}$. In the above equation the "tilde" over  the transverse  indices $k,l$ means they do not include the Killing direction $y$ which is now a direction normal to the D$_{p-1}$-brane. The contracted indices of the second  and  third terms are not complete, \ie the second term has  $y$ which does not include all other transverse coordinates,  and the last term has the index $\tl$ which does not include the transverse coordinate $y$. This indicates that the original action \reef{Tf02} is not consistent with the linear T-duality.

To remedy this failure, one has to add some new couplings. These couplings must be such that when they  combine with the couplings \reef{Tf02}, the indices in the combination must remain complete after  T-duality. Consider  then the following couplings on the world volume of the D$_p$-brane:
\beqa
\frac{T_p}{2!(p-3)!}\int d^{p+1}x\epsilon^{a_0\cdots a_{p-4}abcd}{\cal C}^{(p-3)}_{a_0\cdots a_{p-4}}\bigg[-B_a{}^k{}_{,b}{}^{e}B_{ck,de}+B_a{}^e{}_{,b}{}^{k}B_{ce,dk}\bigg]\labell{Tf3}
\eeqa
Doing the same steps as we have done for the couplings \reef{Tf02}, one finds that the above couplings transforms  to the following couplings of D$_{p-1}$-brane:
\beqa
\frac{T_{p-1}}{2!(p-4)!}\int d^{p}x\epsilon^{a_0\cdots a_{p-5}abcd}{\cal C}^{(p-4)}_{a_0\cdots a_{p-5}}\bigg[-B_a{}^{\tk}{}_{,b}{}^{e}B_{c\tk,d\te}+B_a{}^{e}{}_{,b}{}^{k}B_{ce,dk}+h_a{}^{y}{}_{,b}{}^{k}h_{cy,dk}\bigg]\labell{Tf32}
\eeqa
In this equation also the  index $\tk$ in the first and the index $y$ in the last terms are not complete. This indicates that the coupling \reef{Tf3} is not consistent with the T-duality either. However, the sum of the first term above and the second term of \reef{first00}, and the sum of the last terms above and the last term of \reef{first00} have  complete indices. Hence, the combination of actions \reef{Tf02} and \reef{Tf3}  are consistent with T-duality when $y$ is an index on the R-R potential. That is, 
 the following couplings of D$_p$-brane: 
\beqa
\frac{T_p}{2!(p-3)!}\int d^{p+1}x\epsilon^{a_0\cdots a_{p-4}abcd}{\cal C}^{(p-3)}_{a_0\cdots a_{p-4}}\bigg[h_{a}{}^f{}_{,b}{}^e(h_{ce,df}-h_{cf,de})-B_a{}^k{}_{,b}{}^{e}B_{ck,de}\nonumber\\
-h_{a}{}^l{}_{,b}{}^k(h_{ck,dl}-h_{cl,dk})+B_a{}^e{}_{,b}{}^{k}B_{ce,dk}\bigg]\labell{Tf41}
\eeqa 
 are  consistent with the linear T-duality transformations \reef{linear} when the R-R potential  carries the  Killing index. 

In order to proceed further, one observes that in the actions \reef{Tf41}, two indices $a$ and $c$, which are carried by the metric/B-field terms, contract with the volume form.  When performing T-duality along a particular world volume direction, either one of these or one of the indices on the R-R potential  must equal the T-dual coordinate $y$. We have already shown that  the case, in which the index $y$ is carried by the R-R  field, is consistent with T-duality. Now we  will  check the second case where index $y$ is carried by the metric/B-field terms. The strategy is to choose one of the two indices to perform the T-duality  and infer what extra terms must be included for the consistency. The resulting terms will have one remaining index.  So we repeat  this procedure  to arrive at an action  in which the metric/B-field terms have  no index contracted with the volume form. 

There are two  ways for the metric/B-field terms in \reef{Tf41} to carry the Killing coordinate $y$, \ie either $a$ or $c$ carries the index $y$.  One can write the D$_p$-brane couplings \reef{Tf41} as
\beqa
\frac{T_p}{(p-3)!}\int d^{p+1}x\epsilon^{a_0\cdots a_{p-4}ab yd}{\cal C}^{(p-3)}_{a_0\cdots a_{p-4}}\bigg[h_{a}{}^f{}_{,b}{}^e(h_{ye,df}-h_{yf,de})-B_a{}^k{}_{,b}{}^eB_{yk,de}\nonumber\\
-h_{a}{}^l{}_{,b}{}^k(h_{yk,dl}-h_{yl,dk})+B_a{}^e{}_{,b}{}^kB_{ye,dk}\bigg]\nonumber
\eeqa
Mimicking the steps which are used to get \reef{22}, one finds that  the transformation of the above couplings under T-duality \reef{linear} gives the following couplings for D$_{p-1}$-brane: 
\beqa
\frac{T_{p-1}}{(p-3)!}\int d^{p}x\epsilon^{a_0\cdots a_{p-4}abd}{\cal C}^{(p-2)}_{a_0\cdots a_{p-4}}{}^y\bigg[-h_a{}^f{}_{,b}{}^e(B_{ye,df}-B_{yf,de})+B_a{}^k{}_{,b}{}^e h_{yk,de}\nonumber\\ 
+h_a{}^{l}{}_{,b}{}^{k}(B_{yk,dl}-B_{yl,dk})-B_a{}^e{}_{,b}{}^k h_{ye,dk}\bigg]\labell{y}
\eeqa
In this case the world volume indices $e,f$ and the transverse indices $k,l$ are all complete. However, the $y$ index is not a complete index.  Inspired by the above couplings, one can guess that for the D$_{p}$-brane, the couplings should be following:
\beqa
\frac{T_{p}}{(p-2)!}\int d^{p+1}x\epsilon^{a_0\cdots a_{p-3}abd}{\cal C}^{(p-1)}_{a_0\cdots a_{p-3}}{}^i\bigg[-h_a{}^f{}_{,b}{}^e(B_{ie,df}-B_{if,de})+B_a{}^k{}_{,b}{}^e h_{ik,de}\nonumber\\ 
+h_a{}^{l}{}_{,b}{}^{k}(B_{ik,dl}-B_{il,dk})-B_a{}^e{}_{,b}{}^k h_{ie,dk}\bigg]\labell{new2}
\eeqa
One can easily verify that the above couplings are invariant under the linear T-duality transformations \reef{linear} when the world volume Killing coordinate $y$ is carried by the R-R potential, \ie the R-R potential ${\cal C}^{(p-1)}_{a_0\cdots a_{p-4}y}{}^i$ transforms to ${\cal C}^{(p-2)}_{a_0\cdots a_{p-4}}{}^i$ in which the transverse index $i$ does not include $y$, and the couplings for $i=y$ are given by \reef{y}.

Finally, one observes that there is one  possibility for the metric/B-field terms in \reef{new2} to carry the Killing  coordinate $y$, \ie $a$ carries the index $y$. One can write it as
\beqa
\frac{T_{p}}{(p-2)!}\int d^{p+1}x\epsilon^{a_0\cdots a_{p-3}ybd}{\cal C}^{(p-1)}_{a_0\cdots a_{p-3}}{}^i\bigg[-h_y{}^f{}_{,b}{}^e(B_{ie,df}-B_{if,de})+B_y{}^k{}_{,b}{}^e h_{ik,de}\nonumber\\ 
+h_y{}^{l}{}_{,b}{}^{k}(B_{ik,dl}-B_{il,dk})-B_y{}^e{}_{,b}{}^k h_{ie,dk}\bigg]\nonumber
\eeqa
Under T-duality it transforms to the following couplings for D$_{p-1}$-brane:
\beqa
\frac{T_{p-1}}{(p-2)!}\int d^{p}x\epsilon^{a_0\cdots a_{p-3}bd}{\cal C}^{(p)}_{a_0\cdots a_{p-3}}{}^{iy}\bigg[B_y{}^f{}_{,b}{}^e(B_{ie,df}-B_{if,de})-h_y{}^k{}_{,b}{}^e h_{ik,de}\nonumber\\ 
-B_y{}^{l}{}_{,b}{}^{k}(B_{ik,dl}-B_{il,dk})+h_y{}^e{}_{,b}{}^k h_{ie,dk}\bigg]\labell{yy}
\eeqa
where again,  the world volume indices $e,f$ and the transverse indices $k,l$ are all complete, but $y$ is not. Equation \reef{yy} suggests the following couplings  for the D$_{p}$-brane:
\beqa
\frac{T_{p}}{2!(p-1)!}\int d^{p+1}x\epsilon^{a_0\cdots a_{p-2}bd}{\cal C}^{(p+1)}_{a_0\cdots a_{p-2}}{}^{ij}\bigg[B_j{}^f{}_{,b}{}^e(B_{ie,df}-B_{if,de})-h_j{}^k{}_{,b}{}^eh_{ik,de}\nonumber\\ 
-B_j{}^{l}{}_{,b}{}^{k}(B_{ik,dl}-B_{il,dk})+h_j{}^e{}_{,b}{}^k h_{ie,dk}\bigg]\labell{new31}
\eeqa
One can again verify that the above couplings are invariant under T-duality  when  $y$ is carried by the R-R potential, \ie The R-R potential ${\cal C}^{(p+1)}_{a_0\cdots a_{p-3}y}{}^{ij}$ transforms to ${\cal C}^{(p)}_{a_0\cdots a_{p-3}}{}^{ij}$ in which the transverse indices $i,j$ do not include $y$, and the couplings for $i=y$ or $j=y$ are given by \reef{yy}.

There is no index in the B-field/metric in \reef{new31} that contracts with the volume form. Hence, the combination of couplings \reef{Tf41}, \reef{new2} and \reef{new31} forms a  complete   T-dual multiplet, \ie  the CS multiplet. This multiplet is 
\beqa
T_p\int d^{p+1}x\left(\frac{\epsilon^{a_0\cdots a_{p-4}abcd}}{2!(p-3)!}{\cal C}^{(p-3)}_{a_0\cdots a_{p-4}}\big[h_{a}{}^f{}_{,b}{}^e(h_{ce,df}-h_{cf,de})
%-h_{a}{}^l{}_{,b}{}^k(h_{ck,dl}-h_{cl,dk})\right.\nonumber\\
-B_a{}^k{}_{,b}{}^{e}B_{ck,de}%+B_a{}^e{}_{,b}{}^{k}B_{ce,dk}
\big]\right.\nonumber\\
+\frac{\epsilon^{a_0\cdots a_{p-3}abd}}{(p-2)!}{\cal C}^{(p-1)}_{a_0\cdots a_{p-3}}{}^i\big[-h_a{}^f{}_{,b}{}^e(B_{ie,df}-B_{if,de})%+h_a{}^{l}{}_{,b}{}^{k}(B_{ik,dl}-B_{il,dk})\nonumber\\ 
+B_a{}^k{}_{,b}{}^e h_{ik,de}%-B_a{}^e{}_{,b}{}^k h_{ie,dk}
\big]\nonumber\\
+\frac{\epsilon^{a_0\cdots a_{p-2}bd}}{2!(p-1)!}{\cal C}^{(p+1)}_{a_0\cdots a_{p-2}}{}^{ij}\big[B_j{}^f{}_{,b}{}^e(B_{ie,df}-B_{if,de})%-B_j{}^{l}{}_{,b}{}^{k}(B_{ik,dl}-B_{il,dk})\nonumber\\ 
\left.-h_j{}^k{}_{,b}{}^eh_{ik,de}%+h_j{}^e{}_{,b}{}^k h_{ie,dk}
\big]\frac{}{}\right)-(\cdots)\labell{41}
\eeqa 
where dots refer to the similar expressions as above with  the replacement of the world volume indices $(e,f)$ by the transverse indices $(k,l)$ and  $(e,k)$ by $(k,e)$. We  call the ${\cal C}^{(p-3)}$ couplings  the first component of the multiplet, the ${\cal C}^{(p-1)}$ couplings are called the second  component and so on.  The above couplings have been also found in \cite{Becker:2010ij} and verified with some of the contact terms of the S-matrix element of one R-R and two NS-NS vertex operators. A more details study of the S-matrix element \cite{Garousi:2011ut}, however,  reveals  that the string amplitude has more contact terms than those considered in \cite{Becker:2010ij}.

\subsection{${\cal C}^{(p-3)}$ couplings}

One can easily check that  the first component of the CS multiplet  \reef{41}  is not invariant under the $B$-field gauge  transformations.  
%Using the fact that the S-matrix elements must satisfy the Ward identity corresponding to the B-field gauge transformation,   the fact that the S-matrix element of one RR and two NSNS vertex operators has both contact terms as well as massless poles at order $O(\alpha'^2)$ \cite{Garousi:2011ut}, and the fact that the massless poles at order $O(\alpha'^2)$ which are produced by the supergravity action and the brane couplings \reef{LTdual} are gauge invariant, one realizes that the contact terms must be also invariant under the B-field gauge transformations. 
To write it in terms of field strength $H$, we  add  another T-dual multiplet  to \reef{41}.  Since the gravity couplings to ${\cal C}^{(p-3)}$ are those given by \reef{41} \cite{Craps:1998fn}, the first component of the new T-dual multiplet must include only the B-field. This happens when the  indices of the   R-R potential and the   B-fields    contract either with the volume form or with the derivative of these fields.  Consider the following couplings for ${\cal C}^{(p-3)}$:
\beqa
\frac{T_p}{2!(p-3)!}\int d^{p+1}x\epsilon^{a_0\cdots a_{p-4}abcd}{\cal C}^{(p-3)}_{a_0\cdots a_{p-4}}(B_{ak,be}-B_{ae,bk})B_{cd}{}^{,ek}\labell{new1}
\eeqa 
As indices  $e$ and $k$  appear in  derivatives,   it is  easy to verify that this coupling is invariant under linear T-duality transformations \reef{linear} when the Killing coordinate $y$ is carried by the R-R potential.  When $y$ is carried by the B-field, it is not invariant under T-duality. In those cases one has to add more terms involving the higher R-R forms to arrive at a complete T-dual multiplet.  Applying  the steps that are used  in the previous section, one finds the following T-dual multiplet corresponding to \reef{new1}  
\beqa
&&T_p\int d^{p+1}x\left(\frac{\epsilon^{a_0\cdots a_{p-4}abcd}}{2!(p-3)!}{\cal C}^{(p-3)}_{a_0\cdots a_{p-4}}B_{ak,be}B_{cd}{}^{,ek}\right.\nonumber\\
&&\qquad\qquad\quad+\frac{\epsilon^{a_0\cdots a_{p-3}abd}}{2!(p-2)!}{\cal C}^{(p-1)}_{a_0\cdots a_{p-3}}{}^i\big[h_{ik,be}B_{ad}{}^{,ek}-2B_{ak,be}h_{id}{}^{,ek}\big]\nonumber\\
&&\qquad\qquad\quad+\frac{\epsilon^{a_0\cdots a_{p-2}bd}}{(p-1)!}{\cal C}^{(p+1)}_{a_0\cdots a_{p-2}}{}^{ij}\big[-h_{ik,be}h_{jd}{}^{,ek}
+\frac{1}{2}B_{dk,be}B_{ij}{}^{,ek}\big]\nonumber\\
&&\left.\qquad\qquad\quad+\frac{\epsilon^{a_0\cdots a_{p-1}b}}{2!p!}{\cal C}^{(p+3)}_{a_0\cdots a_{p-1}}{}^{ijn}h_{ie,bk}B_{jn}{}^{,ek}\right)-(\cdots)\labell{7}
\eeqa 
where dots refer to the  expressions similar to the one in the first bracket with indices $(e,k)$ replaced by $(k,e)$. 

Now, the first components of  the CS multiplet \reef{41} and the above multiplet can be written  in terms of $H$, \ie
\beqa
&&\frac{T_p}{2!2!(p-3)!}\int d^{p+1}x\epsilon^{a_0\cdots a_{p-4}abcd}{\cal C}^{p-3}_{a_0\cdots a_{p-4}}\big[\cR_{ab}{}^{ef}\cR_{cdfe}-\cR_{ab}{}^{kl}\cR_{cdlk}\nonumber\\
&&\qquad\qquad\qquad\qquad\qquad\qquad\qquad\qquad-\frac{1}{2}H_{abk,e}H_{cd}{}^{k,e}+\frac{1}{2}H_{abe,k}H_{cd}{}^{e,k}
\big]\labell{Tf41new}
\eeqa
 where $H_{\mu\nu\rho}=B_{\mu\nu,\rho}+B_{\rho\mu,\nu}+B_{\nu\rho,\mu}$.  The terms in the second line are reproduced by the  S-matrix calculation \cite{Garousi:2011ut}. 
 Unlike the gravity couplings in the first line, the  B-field couplings  are not invariant under the R-R gauge transformation.

One may then expect that there are some other  T-dual multiplets  that should be included in the action \reef{Tf41new}.  As we have pointed out above, their  first component must include only the B-field.  %The following are some  possibilities: 
%\beqa
%\int d^{p+1}x\epsilon^{a_0\cdots a_{p-4}abcd}{\cal C}^{(p-3)}_{a_0\cdots a_{p-4}}H_{ab}{}^{\mu,\nu}H_{cd\nu,\mu}+\cdots\,\nonumber\\
%\int d^{p+1}x\epsilon^{a_0\cdots a_{p-4}abcd}{\cal C}^{(p-3)}_{a_0\cdots a_{p-4},\mu}H_{ab\nu}H_{cd}{}^{\mu,\nu}+\cdots\labell{extra}\\
%\int d^{p+1}x\epsilon^{a_0\cdots a_{p-4}abcd}{\cal C}^{(p-3)}_{a_0\cdots a_{p-4},\mu\nu}H_{ab}{}^{\mu}H_{cd}{}^{\nu}+\cdots\nonumber
%\eeqa
%None of  these terms is  invariant under the R-R gauge transformation. However, one may find a combination of them to be gauge invariant. 
%where dots refer to the other components of the multiplets.    
The presence of such couplings can  be fixed by the S-matrix calculation. In fact the couplings \reef{Tf41new} as well as the following couplings are produced by the S-matrix element of one R-R and two NS-NS vertex operators \cite{Garousi:2011ut}:
\beqa
&&\frac{T_{p}}{(p-3)!}\int d^{p+1}x\,\eps^{a_0a_1\cdots a_{p}}{\cal C}^{(p-3)}_{a_4\cdots a_{p}}\left(\frac{1}{2!2!}H^{ea_0a_1}{}_{,ef}H^{fa_2a_3}+\frac{1}{3!}H^{a_0a_1a_2}{}_{,ek}H^{kea_3}\right.\labell{5}\\
&&\qquad\qquad\qquad\qquad\qquad\left.+\frac{1}{2!2!}H^{a_0a_1e,k}{}_eH^{a_2a_3}{}_k+\frac{1}{3!}H^{a_0a_1a_2}{}_{,e}H^{efa_3}{}_{,f}+\frac{1}{3!}H^{a_0a_1a_2}{}_{,k}H^{kea_3}{}_{,e}\right)\nonumber
\eeqa
The S-matrix element also produces  some massless open-string/closed-string poles at order $O(\alpha'^2)$. The open string poles are reproduced in field theory by the DBI action \reef{DBI}  and the following couplings \cite{Garousi:2011ut}:  
\beqa
&&\frac{T_{p}}{(p-3)!}\int d^{p+1}x\,\eps^{a_0a_1\cdots a_{p}}\left(\frac{1}{2!2!}{ C}^{(p-3)}_{a_4\cdots a_p,k}(2H_{a_0a_1}{}^{e,k}{}_e-H_{a_0a_1}{}^{k,e}{}_e)(B_{a_2a_3}+2\pi\alpha'f_{a_2a_3})\right.\nonumber\\
&&\left.-{\cal C}^{(p-3)}_{a_4\cdots a_{p}}\big[\frac{1}{3!} H^{a_0a_1a_2,e}{}_{ef}(B^{fa_3}+2\pi\alpha' f^{fa_3})+\frac{1}{2!2!} H^{a_0a_1f,e}{}_{e}(B^{a_2a_3}+2\pi\alpha' f^{a_2a_3}){}_{,f}\big]
\right)\labell{6}
\eeqa
The closed string poles, on the other hand,  can be reproduced by the bulk supergravity and the D-brane couplings \reef{LTdual}. It is shown in \cite{Garousi:2011ut}  that even though the contact terms and the massless poles   are not separately invariant under the R-R gauge transformations, their combination satisfies this symmetry.

%One may suspect that the complete CS action at order $O(\alpha'^2)$ may have  couplings like ${\cal C}^{(p-3)}H^4$. However, this term is not invariant under T-duality. One needs to include terms like ${\cal C}^{(p-3)}(\prt h)^4$ to have  T-duality invariant couplings. Such gravity couplings can not be written in covariant form, and are not reproduced by the S-matrix calculations \cite{Craps:1998fn}.

\subsection{${\cal C}^{(p-1)}$ couplings}

Before making  the other components  to be covariant/gauge-invariant, let us digress to discuss a  subtle point in finding the field theory couplings from the corresponding string theory  S-matrix elements. It has been shown in \cite{Garousi:2011ut} that the S-matrix element of one R-R potential $C^{(p+5)}$ and two gravitons is zero, and the S-matrix element of one $C^{(p+5)}$ and two B-fields is non-zero. When writing the latter amplitude in terms of field strength $H$, one finds that it has only massless closed string poles at order $O(\alpha'^2) $ (see  appendix B in \cite{Garousi:2011ut}). On the other hand, using the  steps that are applied in  section 3.1, one finds that the T-dual multiplet corresponding to the couplings \reef{Tf41new} has no $C^{(p+5)}$ component. However, for the couplings \reef{5}, one find following component in the T-dual multiplet:
\beqa
C^{(p+5)}_{a_0\cdots a_p}{}^{ijmn}B^{ij,e}{}_{ef}B^{mn,k}+C^{(p+5)}_{a_0\cdots a_p}{}^{ijmn}B^{ij,e}{}_{ek}B^{mn,k}\labell{10}
\eeqa
which arises from the contact terms of the S-matrix element.  Similarly for the couplings \reef{6}, T-dual multiplet has following component:
 \beqa
C^{(p+5)}_{a_0\cdots a_p}{}^{ijmn,k}B^{ij,e}{}_{ek}B^{mn}-C^{(p+5)}_{a_0\cdots a_p}{}^{ijmn}B^{ij,e}{}_{ef}B^{mn,f}\labell{11}
\eeqa
which arises from the massless open string  poles of the S-matrix element. The S-matrix element also produces some massless closed string poles.   

The above results indicate  that the  contact terms \reef{10} and \reef{11} must be canceled by the  massless closed string poles. To see this explicitly,  we apply the same steps as we have used in  section 3.1 to calculate the $C^{(p+5)}$ component of the S-matrix element of one $C^{(p-3)}$ and two B-fields \cite{Garousi:2011ut}. The $C^{(p+5)}$ component of the amplitude (35) in  \cite{Garousi:2011ut} is
\beqa
{\cal A}&\sim& p_2\inn V\inn p_2[p_3\inn V\inn p_3\cJ_3-\frac{1}{2}p_2\inn V\inn p_3\cJ_1+\frac{1}{2}p_2\inn N\inn p_3\cJ_2-p_1\inn N\inn p_3\cI_7]\nonumber\\
&&+\frac{1}{4}p_1\inn N\inn p_2[p_1\inn N\inn p_3\cI_1-p_2\inn N\inn p_3\cI_2+p_2\inn \inn p_3\cI_3]+(2\leftrightarrow 3)
\eeqa
The amplitude  also has the overall factor of the polarization of the external states, \ie $\eps^{a_0\cdots a_p}{\veps _1}_{a_0\cdots a_p i j mn}\veps_2^{ij}\veps_3^{mn}$. In the above equation, $\cJ$s and $\cI$s are  functions of the Mandelstam variables. We refer the interested reader to \cite{Garousi:2011ut} for the notations. At low energy, the terms in the first line produce the contact terms \reef{10} and \reef{11} and some massless closed string poles, whereas the terms in the second line produce only massless closed string poles. Using the identities between   $\cJ$s and $\cI$s, \ie eq.(33) in \cite{Garousi:2011ut}, one finds that the terms in the first line add up to zero, as anticipated above. On the other hand, the terms in the second line combine with some  other S-matrix T-dual multiplets to produce the result for the S-matrix element of one $C^{(p+5)}$ and two B-fields.

This phenomenon may happen only when two derivatives in a coupling contract with each other or, in momentum space, when a contact term is proportional to a Mandelstam variable, \eg the couplings in \reef{10} and \reef{11}.   The reason  is that the identities between   $\cJ$s and $\cI$s which arise from the requirement that the S-matrix element must satisfy the Ward identities corresponding to the NS-NS massless fields, have the structure of $\sum_i (M_i\cJ_i+N_i\cI_i)=0$ where $M_i$ and $N_i$ are at least the linear order of some Mandelstam variables. To clarify this point, suppose the S-matrix element has the structure $\sum_i(\cdots)\cJ_i$ where $(\cdots)$ refers to the polarization tensors and the four momenta that are produced by performing the correlation between the vertex operators \cite{Garousi:2010bm}. Upon imposing the Ward identity corresponding to one of the NS-NS states, \ie $\veps_{\mu\nu}\rightarrow \veps_{\mu\nu}+p_{\mu}\z_{\nu}\pm p_{\nu}\z_{\mu}$ one finds a relation like $\sum_iM_i\cJ_i=0$. 

These identities make  the covariant/gauge-invariant form of  the couplings, in which two derivatives contract with each other, ambiguous. This ambiguity, however, is resolved when one considers both the contact terms as well as the massless poles at order $O(\alpha'^2)$.  
  Hence, we will not  discuss the covariant/gauge-invariant form of such couplings, \eg we will not consider the $C^{(p-1)}$, $C^{(p+1)}$ and $C^{(p+3)}$ components of \reef{7}. Note that  the first component of this multiplet is combined with the first component of the CS multiplet to produce the gauge invariant result \reef{Tf41new}.

We now try to make the other components of the CS multiplet to be covariant/gauge-invariant. One can easily verify that the structure of the couplings in \reef{5} and \reef{6} is different from the structure of the couplings in the CS multiplet \reef{41}. In particular, the couplings \reef{41} are antisymmetric under $(e,f)\rightarrow (k,e)$ and $(e,k)\rightarrow (k,l)$ whereas  the couplings \reef{5} and \reef{6} do not have such antisymmetric property. So the couplings \reef{41}  can not be combined with the T-dual multiplets corresponding to the couplings \reef{5} and \reef{6} to produce the covariant/gauge-invariant results. Therefore, there must be other T-dual multiplets to make the CS multiplet covariant/gauge-invariant. The first component of these multiplets should be ${\cal C}^{(p-1)}$.  The strategy for finding these multiplets is to find its first component by requiring that when they are  combined  with the corresponding coupling in \reef{41}, they become  covariant/gauge-invariant. Then using the same steps as in section 3.1, one finds all the other components of the T-dual multiplets. 

There are two multiplets for making  the first term in the second line of the CS multiplet \reef{41} to be invariant under the B-field gauge transformations.  The first multiplet which has only two components, is given by the following expression: 
\beqa
&\!\!\!\!\!\!\!&\alpha T_p\int d^{p+1}x\left(\frac{\epsilon^{a_0\cdots a_{p-3}abd}}{(p-2)!}{\cal C}^{(p-1)}_{a_0\cdots a_{p-3}}{}^i\big[-h_a{}^f{}_{,b}{}^eB_{ef,id}-B_a{}^k{}_{,b}{}^eh_{ek,id}\big]\nonumber\right.\\
&\!\!\!\!\!\!\!&\qquad\qquad\left.+\frac{\epsilon^{a_0\cdots a_{p-2}bd}}{(p-1)!}{\cal C}^{(p+1)}_{a_0\cdots a_{p-2}}{}^{ij}\big[B_j{}^f{}_{,b}{}^eB_{ef,id}+h_j{}^k{}_{,b}{}^eh_{ek,id}\big]\frac{}{}\right)-(\cdots)\labell{new22}
\eeqa
where again dots refer to the  similar terms as above with  $(e,f)\rightarrow (k,l)$ and  $(e,k)\rightarrow (k,e)$.  The coefficient $\alpha$ is a constant which we will determine  shortly. Note that the first  term in the first line above is the coupling that is needed to make the corresponding coupling in \reef{41}   gauge invariant. The other term in the first line is needed for T-duality. One can easily check that the sum of the first term above for $\alpha=1$ and the first term in the second line of \reef{41} can be written in terms of $H$, \ie $\cR_{ab}{}^{ef}H_{ief,d}$. However, the sum of the first term in the second line above for $\alpha=1$ and the first term in the third line of \reef{41} can not be written in a gauge invariant form.
% if one adds another multiplet which includes ${\cal C}^{(p+1)}_{a_0\cdots a_{p-2}}{}^{ij}(B^{fe}{}_{,jb}B_{ef,id}
%-B^{lk}{}_{,jb}B_{kl,id}+\cdots)$. This indicates that the S-matrix element of one R-R and two NSNS vertex operators must have a contact term like $p_2^ip_3^jp_2^{b}p_2^{d}\Tr(\veps_2\inn D\inn\veps_3)$ where $\veps_2$ and $\veps_3$ are the polarization of the B-fields. Using the general form of the S-matrix element in \cite{...}, one can easily verify that such structure can not be reproduced by the S-matrix element. 

The other  multiplet is:
\beqa
&\!\!\!\!\!\!\!&\beta T_p\int d^{p+1}x\left(\frac{\epsilon^{a_0\cdots a_{p-3}abd}}{(p-2)!}{\cal C}^{(p-1)}_{a_0\cdots a_{p-3}}{}^i\big[-(h_a{}^f{}_{,b}{}^e-h_a{}^e{}_{,b}{}^f)B_{ed,if}-B_a{}^k{}_{,b}{}^eh_{dk,ie}\big]\nonumber\right.\\
&\!\!\!\!\!\!\!&\qquad\qquad+\frac{\epsilon^{a_0\cdots a_{p-2}bd}}{(p-1)!}{\cal C}^{(p+1)}_{a_0\cdots a_{p-2}}{}^{ij}\big[(B_j{}^f{}_{,b}{}^e-B_j{}^e{}_{,b}{}^f)B_{ed,if}-B_d{}^k{}_{,b}{}^eB_{jk,ie}\nonumber\\
&&\qquad\qquad\qquad\qquad\qquad\qquad\qquad-(h_b{}^f{}_{,d}{}^e-h_b{}^e{}_{,d}{}^f)h_{ej,if}+h_j{}^k{}_{,b}{}^eh_{dk,ie}\big]\nonumber\\
&&\qquad\qquad\left.+\frac{\epsilon^{a_0\cdots a_{p-1}b}}{p!}{\cal C}^{(p+3)}_{a_0\cdots a_{p-1}}{}^{ijn}\big[(B_j{}^f{}_{,b}{}^e-B_j{}^e{}_{,b}{}^f)h_{en,if}+h_n{}^k{}_{,b}{}^eB_{jk,ie}\big]\frac{}{}\right)-(\cdots)\labell{new223}
\eeqa
 The coefficient $\beta$ is a constant. One can  check that the sum of the first term above for $\beta=1$ and the first term in the second line of \reef{41} can also be written in terms of $H$, \ie $\cR_{ab}{}^{ef}H_{ide,f}$. The sum of the first term in the second line above for $\beta=1$ and the first term in the third line of \reef{41} can not be written in a gauge invariant form. 
%by adding a new multiplet whose first component starts at $C^{p+1}$. 
To remedy these failures, we will consider both multiplets with 
\beqa
\alpha=\beta=1/2
\eeqa
We will see in the next section that the above choice of the constants makes it possible to write the first term in the third line of \reef{41} in a gauge invariant form.

The last term in the first line of \reef{new223}  is proportional to the Mandelstam variable $p_2\inn V\inn p_3$ in the momentum space. So as argued before, we are not interested in making it  covariant/gauge-invariant. However, the last term in the first line of \reef{new22} is not proportional to a Mandelstam variable, so there must be other T-dual multiplets to make it covariant/gauge-invariant. One can write it in covariant form by adding the terms $B_a{}^k{}_{,b}{}^e(h_{id,ek}-h_{ik,de}-h_{de,ik})$. The first two terms  are again proportional to the  Mandelstam variable $p_2\inn V\inn p_3$ and are not relevant for our discussion here. The last term is the $C^{(p-1)}$ component of the following multiplet: 
\beqa
&\!\!\!\!\!\!\!&\alpha T_p\int d^{p+1}x\left(\frac{\epsilon^{a_0\cdots a_{p-3}abd}}{(p-2)!}{\cal C}^{(p-1)}_{a_0\cdots a_{p-3}}{}^i\big[B_a{}^k{}_{,b}{}^eh_{de,ik}\big]\nonumber\right.\\
&\!\!\!\!\!\!\!&\qquad\qquad+\frac{\epsilon^{a_0\cdots a_{p-2}bd}}{(p-1)!}{\cal C}^{(p+1)}_{a_0\cdots a_{p-2}}{}^{ij}\big[-h_j{}^k{}_{,b}{}^eh_{de,ik}
\big]\nonumber\\
&&\qquad\qquad\left.+\frac{\epsilon^{a_0\cdots a_{p-1}b}}{p!}{\cal C}^{(p+3)}_{a_0\cdots a_{p-1}}{}^{ijn}\big[h_j{}^k{}_{,b}{}^eB_{ne,ik}\big]\frac{}{}\right)-(\cdots)\labell{new224}
\eeqa
%Again, there are similar expressions as above with opposite sign and with replacing  $(e,k)$ by $(k,e)$.

Now, the ${\cal C}^{(p-1)}$ component of  the above multiplet and  the multiplets \reef{new223}, \reef{new22}  and  \reef{41} add up to the following covariant/gauge-invariant results:
\beqa
&&\frac{T_{p}}{(p-2)!}\int d^{p+1}x\epsilon^{a_0\cdots a_{p-3}abd}{\cal C}^{(p-1)}_{a_0\cdots a_{p-3}}{}^i\bigg[-\frac{1}{2}\cR_{ab}{}^{ef}(H_{ied,f}+\frac{1}{2}H_{ief,d})-\cR_{iekd}H_{ab}{}^{k,e}\nonumber\\ 
&&\qquad\qquad\qquad\qquad\qquad\qquad\qquad\qquad+\frac{1}{2}\cR_{ab}{}^{kl}(H_{ikd,l}+\frac{1}{2}H_{ikl,d})+\cR_{iked}H_{ab}{}^{e,k}\bigg]\labell{new24}
\eeqa
 where we have added/removed some terms which are proportional to the Mandelstam variables. They are related to the massless-pole T-dual multiplets and we are not concerned about them  in this paper.

\subsection{${\cal C}^{(p+1)}$ couplings}

Adding the multiplets \reef{new22} and \reef{new223} to the CS multiplet \reef{41}, one finds that the B-field terms in the ${\cal C}^{(p+1)}$ component of the CS multiplet can  be written in terms of field strength $H$, provided  one adds  one more  multiplet, \ie
\beqa
&\!\!\!\!\!\!\!&\frac{1}{2} T_p\int d^{p+1}x\left(\frac{\epsilon^{a_0\cdots a_{p-2}bd}}{(p-1)!}{\cal C}^{(p+1)}_{a_0\cdots a_{p-2}}{}^{ij}\big[B^{fe}{}_{,jb}B_{ed,if}-h^{ek}{}_{,jb}h_{kd,ie}
\big]\nonumber\right.\\
&&\qquad\qquad\left.+\frac{\epsilon^{a_0\cdots a_{p-1}b}}{p!}{\cal C}^{(p+3)}_{a_0\cdots a_{p-1}}{}^{ijn}\big[B^{fe}{}_{,jb}h_{en,if}-h^{ek}{}_{,jb}B_{kn,ie}\big]\frac{}{}\right)-(\cdots)\labell{new225}
\eeqa

Now, the ${\cal C}^{(p+1)}$ components of  the multiplets \reef{41}, \reef{new22}, \reef{new223}, \reef{new224}, and \reef{new225}  add up to the following covariant/gauge-invariant result:
\beqa
&&\frac{T_{p}}{(p-1)!}\int d^{p+1}x\epsilon^{a_0\cdots a_{p-2}bd}{\cal C}^{(p+1)}_{a_0\cdots a_{p-2}}{}^{ij}\bigg[\frac{1}{2}H_j{}^{fe}{}_{,b}H_{ied,f}+\frac{1}{4}\cR_{bd}{}^{ef}\cR_{ijfe}+\cR^e{}_{jb}{}^k\cR_{eidk}\nonumber\\
&&\qquad\qquad\qquad\qquad\qquad\qquad\qquad-\frac{1}{2}H_j{}^{lk}{}_{,b}H_{ikd,l}-\frac{1}{4}\cR_{bd}{}^{kl}\cR_{ijlk}-\cR^k{}_{jb}{}^e\cR_{kide}\bigg]\labell{new25}
\eeqa
where again we have added/removed some terms which are proportional to the Mandelstam variables.  One may also add $H_{bd}{}^{k,e}H_{ijk,e}-H_{bd}{}^{e,k}H_{ije,k}$ to the above bracket. Since this contact term is proportional to the Mandelstam variables, our present calculation, which does not consider the massless poles, can not confirm the presence of this coupling. %This term however is very much line the couplings in the second line of \reef{Tf41new}. So it is very likely that these couplings are also reproduced by the S-matrix calculation.

\subsection{${\cal C}^{(p+3)}$ couplings}

The CS multiplet \reef{41} does not have a $C^{(p+3)}$ component. However, the multiplets   \reef{new223}, \reef{new224} and \reef{new225}, which have made the CS multiplet covariant/gauge-invariant have such a component.  They combined to the following covariant/gauge invariant result:
\beqa
&&\frac{T_{p}}{p!}\int d^{p+1}x\epsilon^{a_0\cdots a_{p-1}b}{\cal C}^{(p+3)}_{a_0\cdots a_{p-1}}{}^{ijn}\bigg[\frac{1}{4}H_j{}^{fe}{}_{,b}\cR_{nife}-\frac{1}{4}H_{ni}{}^{e,k}\cR_{jebk}\nonumber\\
&&\qquad\qquad\qquad\qquad\qquad\qquad-\frac{1}{4}H_j{}^{kl}{}_{,b}\cR_{nikl}+\frac{1}{4}H_{ni}{}^{k,e}\cR_{jkbe}\bigg]\labell{new26}
\eeqa
There is no  coupling for $C^{(p+5)}$ which is consistent with the S-matrix calculation \cite{Garousi:2011ut}.
 
Therefore the couplings \reef{Tf41new}, \reef{new24}, \reef{new25} and \reef{new26} are the T-dual multiplet corresponding to the CS multiplet \reef{41} which are covariant and are invariant under the B-field gauge transformations. The  T-duality of the multiplet, however, is off by  some  contact terms which are proportional to the Mandelstam variables. As we argued in section 3.3, they are related to the massless poles. This ends our construction of making the CS multiplet \reef{41} covariant/gauge-invariant by adding new T-dual multiplets. One can use the same technique as we have done in this paper to find the covariant/gauge-invariant T-dual multiplets corresponding to the couplings \reef{5} and \reef{6}, and then confirm  the results with the S-matrix calculation.

Our studies   indicate that the object that must be invariant under the T-duality is the S-matrix element, not the low energy field theory of the D-brane.  For the cases where the S-matrix element has only contact terms at a given order of $\alpha'$, the field theory is invariant under T-duality at that order. In other cases, the combination of the D-brane contact terms and the massless poles arising from the bulk and the brane actions, is invariant under T-duality. 
The same thing is true for the R-R gauge transformation of the D-brane action at order $O(\alpha'^2)$.  One can check that the couplings \reef{Tf41new} and \reef{5} are not invariant under the R-R gauge transformations. However, taking  the transformation of the massless poles at order $O(\alpha'^2)$ into account, one recovers the R-R gauge symmetry \cite{Garousi:2011ut}.

{\bf Acknowledgments}: I would like to thank Katrin Becker and Rob Myers for   discussions and collaboration in the early stage of this work. I would also like to thank  Ajay Singh for editing the manuscript.
  This work is supported by Ferdowsi University of Mashhad under grant 2/15298-1389/07/11.


\begin{thebibliography}{10}



%\cite{Leigh:1989jq}
\bibitem{Leigh:1989jq}
  R.~G.~Leigh,
  %``Dirac-Born-Infeld Action from Dirichlet Sigma Model,''
  Mod.\ Phys.\ Lett.\  A {\bf 4}, 2767 (1989).
  %%CITATION = MPLAE,A4,2767;%%



%\cite{Bachas:1995kx}
\bibitem{Bachas:1995kx}
  C.~Bachas,
  %``D-brane dynamics,''
  Phys.\ Lett.\  B {\bf 374}, 37 (1996)
  [arXiv:hep-th/9511043].
  %%CITATION = PHLTA,B374,37;%%
 %\cite{Bachas:1999um}
\bibitem{Bachas:1999um}
  C.~P.~Bachas, P.~Bain and M.~B.~Green,
  %``Curvature terms in D-brane actions and their M-theory origin,''
  JHEP {\bf 9905}, 011 (1999)
  [arXiv:hep-th/9903210].
  %%CITATION = JHEPA,9905,011;%% 
  
%\cite{Garousi:1996ad}
\bibitem{Garousi:1996ad}
  M.~R.~Garousi and R.~C.~Myers,
  %``Superstring Scattering from D-Branes,''
  Nucl.\ Phys.\  B {\bf 475}, 193 (1996)
  [arXiv:hep-th/9603194].
  %%CITATION = NUPHA,B475,193;%%
%\cite{Hashimoto:1996kf}
\bibitem{Hashimoto:1996kf}
  A.~Hashimoto and I.~R.~Klebanov,
  %``Decay of Excited D-branes,''
  Phys.\ Lett.\  B {\bf 381}, 437 (1996)
  [arXiv:hep-th/9604065].
  %%CITATION = PHLTA,B381,437;%%

%\cite{Garousi:2009dj}
\bibitem{Garousi:2009dj}
  M.~R.~Garousi,
  %``T-duality of Curvature terms in D-brane actions,''
  JHEP {\bf 1002}, 002 (2010)
  [arXiv:0911.0255 [hep-th]].
  %%CITATION = JHEPA,1002,002;%%




%\cite{Polchinski:1995mt}
\bibitem{Polchinski:1995mt}
  J.~Polchinski,
  %``Dirichlet-Branes and Ramond-Ramond Charges,''
  Phys.\ Rev.\ Lett.\  {\bf 75}, 4724 (1995)
  [arXiv:hep-th/9510017].
  %%CITATION = PRLTA,75,4724;%%

%\cite{Douglas:1995bn}
\bibitem{Douglas:1995bn}
  M.~R.~Douglas,
  %``Branes within branes,''
  arXiv:hep-th/9512077.
  %%CITATION = HEP-TH/9512077;%%




%\cite{Green:1996dd}
\bibitem{Green:1996dd}
  M.~B.~Green, J.~A.~Harvey and G.~W.~Moore,
  %``I-brane inflow and anomalous couplings on D-branes,''
  Class.\ Quant.\ Grav.\  {\bf 14}, 47 (1997)
  [arXiv:hep-th/9605033].
  %%CITATION = CQGRD,14,47;%%
%\cite{Cheung:1997az}
\bibitem{Cheung:1997az}
  Y.~K.~Cheung and Z.~Yin,
  %``Anomalies, branes, and currents,''
  Nucl.\ Phys.\  B {\bf 517}, 69 (1998)
  [arXiv:hep-th/9710206].
  %%CITATION = NUPHA,B517,69;%%
%\cite{Minasian:1997mm}
\bibitem{Minasian:1997mm}
  R.~Minasian and G.~W.~Moore,
  %``K-theory and Ramond-Ramond charge,''
  JHEP {\bf 9711}, 002 (1997)
  [arXiv:hep-th/9710230].
  %%CITATION = JHEPA,9711,002;%%
%\cite{Garousi:2010bm}
\bibitem{Garousi:2010bm}
  M.~R.~Garousi and M.~Mir,
  %``On RR couplings on D-branes at order $O(\alpha'^2)$,''
  JHEP {\bf 1102}, 008 (2011)
  [arXiv:1012.2747 [hep-th]].
  %%CITATION = JHEPA,1102,008;%%


%\cite{Myers:1999ps}
\bibitem{Myers:1999ps}
  R.~C.~Myers,
  %``Dielectric-branes,''
  JHEP {\bf 9912}, 022 (1999)
  [arXiv:hep-th/9910053].
  %%CITATION = JHEPA,9912,022;%%
  %\cite{Hull:2009mi}
\bibitem{Hull:2009mi}
  C.~Hull and B.~Zwiebach,
  %``Double Field Theory,''
  JHEP {\bf 0909}, 099 (2009)
  [arXiv:0904.4664 [hep-th]].
  %%CITATION = JHEPA,0909,099;%%
%\cite{Becker:2010ij}
\bibitem{Becker:2010ij}
  K.~Becker, G.~Guo and D.~Robbins,
  %``Higher Derivative Brane Couplings from T-Duality,''
  arXiv:1007.0441 [hep-th].
  %%CITATION = ARXIV:1007.0441;%%

%\cite{Garousi:2010ki}
\bibitem{Garousi:2010ki}
  M.~R.~Garousi,
  %``Ramond-Ramond field strength couplings on D-branes,''
  JHEP {\bf 1003}, 126 (2010)
  [arXiv:1002.0903 [hep-th]].
  %%CITATION = JHEPA,1003,126;%%
\bibitem%[AAT]
{AAT}{A.A. Tseytlin, Nucl. Phys. B {\bf 276} (1987) 391.}
\bibitem%[AAT1]
{AAT1}{A.A. Tseytlin, Phys. Lett. B {\bf 176} (1986) 92; R.R.
Metsaev, A.A. Tseytlin, Phys. Lett. B {\bf 185} (1987) 52.}
  
  
  
\bibitem{TB}  
T. Buscher, Phys. Lett. B {\bf 159} (1985) 127; B {\bf 194} (1987) 59; B {\bf 201} (1988) 466.

%\cite{Meessen:1998qm}
\bibitem{Meessen:1998qm}
  P.~Meessen and T.~Ortin,
  %``An Sl(2,Z) multiplet of nine-dimensional type II supergravity theories,''
  Nucl.\ Phys.\  B {\bf 541}, 195 (1999)
  [arXiv:hep-th/9806120].
  %%CITATION = NUPHA,B541,195;%%
  
%\cite{Bergshoeff:1995as}
\bibitem{Bergshoeff:1995as}
  E.~Bergshoeff, C.~M.~Hull and T.~Ortin,
  %``Duality in the type II superstring effective action,''
  Nucl.\ Phys.\  B {\bf 451}, 547 (1995)
  [arXiv:hep-th/9504081].
  %%CITATION = NUPHA,B451,547;%%
%\cite{Bergshoeff:1996ui}
\bibitem{Bergshoeff:1996ui}
  E.~Bergshoeff, M.~de Roo, M.~B.~Green, G.~Papadopoulos and P.~K.~Townsend,
  %``Duality of Type II 7-branes and 8-branes,''
  Nucl.\ Phys.\  B {\bf 470}, 113 (1996)
  [arXiv:hep-th/9601150].
  %%CITATION = NUPHA,B470,113;%%

%\cite{Hassan:1999bv}
\bibitem{Hassan:1999bv}
  S.~F.~Hassan,
  %``T-duality, space-time spinors and R-R fields in curved backgrounds,''   
  Nucl.\ Phys.\  B {\bf 568}, 145 (2000)
  [arXiv:hep-th/9907152].
  %%CITATION = NUPHA,B568,145;%%


  
  %\cite{Taylor:1999pr}
\bibitem{Taylor:1999pr}
  W.~Taylor and M.~Van Raamsdonk,
  %``Multiple Dp-branes in weak background fields,''
  Nucl.\ Phys.\  B {\bf 573}, 703 (2000)
  [arXiv:hep-th/9910052].
  %%CITATION = NUPHA,B573,703;%%
%\cite{Craps:1998fn}
\bibitem{Craps:1998fn}
  B.~Craps and F.~Roose,
  %``Anomalous D-brane and orientifold couplings from the boundary state,''
  Phys.\ Lett.\  B {\bf 445}, 150 (1998)
  [arXiv:hep-th/9808074].
  %%CITATION = PHLTA,B445,150;%%
  
%\cite{Garousi:2011ut}
\bibitem{Garousi:2011ut}
  M.~R.~Garousi and M.~Mir,
  %``Towards extending the Chern-Simons couplings at order $O(\alpha'^2)$,''
  arXiv:1102.5510 [hep-th].
  %%CITATION = ARXIV:1102.5510;%%


%\cite{Hassan:2000zk}
%\bibitem{Hassan:2000zk}
%  S.~F.~Hassan and R.~Minasian,
  %``D-brane couplings, RR fields and Clifford multiplication,''
%  arXiv:hep-th/0008149.
  %%CITATION = HEP-TH/0008149;%%








  

\end{thebibliography}
\end{document}